# About the Infinite Repetition of Histories in Space

## Francisco José Soler Gil[1] and Manuel Alfonseca[2]

[1] Universidad de Sevilla, Spain
[2] Escuela Politécnica Superior, Universidad Autónoma de Madrid, Spain

**Abstract**

**This paper analyzes two different proposals, one by Ellis and Brundrit, based on classical relativistic cosmology, the other by Garriga and Vilenkin, based on the DH interpretation of quantum mechanics, both of which conclude that, in an infinite universe, planets and living beings must be repeated an infinite number of times. We point to some possible shortcomings in the arguments of these authors. We conclude that the idea of an infinite repetition of histories in space cannot be considered strictly speaking a consequence of current physics and cosmology. Such ideas should be seen rather as examples of «ironic science» in the terminology of John Horgan.**

## 1. Introduction

The notion that events and series of events make history happen once and again in an infinite repetition is very old. In the West it is at least as old as philosophy[1] and it can also be found in even older traditions of mythic thought[2].

These repetitions are usually presented as recurrences in time, not in space. This is self-explaining, for it is modern astronomy which has first opened the possibility of considering planet Earth as a non-unique object, as the representative of a class, possibly large, of life-sustaining planets.

Fantastic literature goes usually before speculative science. Thus, the idea that the universe is repetitive (in both spatial and temporal senses), that the same situations repeat once and again, appears first in narrations such as *La biblioteca de Babel*, by Borges[3]. In this story, Borges describes a library, a clear image of the universe, which contains all the possible books, even those which are made of meaningless character strings. Every book contains exactly 1,312,000 characters. For every particular book, the library contains over 30 million almost identical copies which differ in just one character (they can be considered as copies with a single errata); almost one quadrillion copies differing in two characters; and so forth. In addition, the library also contains the translation of every book to all the tongues, actual or just possible. However, both the library and the number of books it contains are not infinite. If the library were infinite (Borges suggests this possibility at the end of the story) each book would be represented by an infinite number of copies.

---

[1] Stoic philosophers maintained this position, apparently taken from Heraclitus, but it is possible it goes back as far as Anaximander.

[2] See Eliade, M. (1954), *The Myth of the Eternal Return: Cosmos and History* (Princeton, Princeton University Press).

Up to now, these ideas could seem appropriate only for mythology and fantastic literature. But recently a new possibility has risen: that they can be derived from physical theories about the universe. Soon after the big-bang cosmology was established as the standard cosmological model, some arguments were offered which, starting from apparently plausible assumptions, seem to bring us to the scenario of a universe of histories which repeat infinitely in space.

The objectives of this paper are: (1) summarize the two main argumentations that end in this conclusion; (2) criticize some possible errors in those argumentations; (3) discuss the controversial philosophical concept implicit in both: the actual possibility of an infinite universe; and (4) signal the difference between the highly conjectural character of the argumentations discussed here and the way in which they are frequently described in popular science magazines and books.

To fulfill these goals, the remainder of this article is divided thus: sections two and three describe the two most representative argumentations among those that lead to a universe with an infinite number of histories that repeat in space: the argumentation by Ellis and Brundrit, and the argumentation by Garriga and Vilenkin. Sections four and five point out the possible shortcomings in those argumentations. The sixth section discusses the controversial character of the infinite universe. And the seventh compares the result of this analysis with the way in which those speculations are presented in the popular science media.

---

[3] Borges, J. L. (1944), «*La biblioteca de Babel*».In: *Ficciones* (Buenos Aires, Sur).

## 2. The Ellis-Brundrit (E-B) argumentation

The argumentation proposed by Ellis and Brundrit[4] starts from classic relativistic physics and the hypothesis that the universe is spatially infinite and homogeneous, and develops as follows:

1. The number of galaxies and planets in an infinite homogeneous universe is infinite.

2. The number of possible history lines in the configuration space of that universe is also infinite.

3. Among those history lines, a few could give room to the emergence of living beings based on DNA on the surface of a planet. The number of line histories compatible with life must be at least one, since we are here.

4. If the probability of DNA-based life should be greater than zero, an infinite universe will contain an infinite number of living beings of this type. To explain why this probability may be greater than zero, they claim the following:

    «We assume that this is a stable property, that is, the probability remains finite [they mean non-zero] for a small range of conditions different from those holding in our own vicinity in the universe (e.g. a small change in average temperature or average matter-density would not reduce this probability to zero»[5].

    What they mean is that different but similar history lines would end in the same or similar results.

5. The number of possible DNA-based living beings is finite, because the size of the DNA molecules cannot be arbitrarily large.

6. Therefore an infinite universe will contain an infinite number of copies of every possible living being. Applying the same reasoning a further step, since the possible material combinations on a planet are also finite, we'll conclude that the universe will contain an infinity of planets identical to Earth, including all the living beings on it.

---

[4] Ellis, G.F.R. - Brundrit, G.B. (1979), «Life in the Infinite Universe» *Quarterly Journal of the Royal Astronomical Society* 20 (1979) 37-41. Available at: http://adsabs.harvard.edu/abs/1979QJRAS..20...37E.

[5] Ellis, G.F.R. - Brundrit, G.B. (1979) 38.

7. Therefore we live in a universe where events and histories are repeated in an infinity of places. The authors explain it thus:

    a. «...with an infinite number of histories to look at, it is difficult to provide an argument why a particular history should occur only once»[6].

    b. «If initial conditions are such that the probability of any event occurring is non-vanishing, then we expect that event to occur in infinitely many planetary systems»[7].

## 3. The Garriga-Vilenkin (G-V) argumentation

In their paper «Many worlds in one»[8] Garriga and Vilenkin propose a different scenario, based on the decoherent histories (DH) interpretation of quantum mechanics[9]. Their argumentation develops in the following way:

1. They assume that the universe (or the inflationary multiverse, which appears to be their preferred option) is spatially infinite and homogeneous. That universe (or multiverse) can be divided into an infinity of causally disconnected regions ($O$-regions) defined by event horizons.

2. They assume that space is quantifiable, so that every $O$-region can be divided in a finite number of cells.

3. In addition, they assume that this quantification is correctly described by the DH interpretation of quantum mechanics.

---

[6] Ellis, G.F.R. - Brundrit, G.B. (1979) 38.

[7] Ellis, G.F.R. - Brundrit, G.B. (1979) 41 (endnote b).

[8] Garriga, J. - Vilenkin, A. (2001), «Many worlds in one». Available at http://arxiv.org/abs/gr-qc/0102010.

[9] Gell-Mann, M. - Hartle, J. (1990), «Quantum mechanics in the light of quantum cosmology». In: Zurek, W. (ed.) (1990), *Complexity, Entropy and the physics of information* SFI studies in the Sciences of Complexity, Vol. VIII (Addison-Wesley, Reading); Gell-Mann, M. - Hartle, J.(1993), «Classical equations for quantum systems» *Phys. Rev*. D. 47, (1993) 3345. Available at arXiv:gr-qc/9210010.

4. If the energy in every $O$-region is finite, since energy is quantified, it would be possible to divide its range of variation into a finite number of bins; therefore the number of possible histories in an $O$-region would be finite. In those environments where planets and life can exist, such as Earth, energy can be considered finite, since we are far from gravity singularities. This means that gravity can be ruled out of this argumentation.

5. Therefore, given that by hypothesis there is an infinite number of $O$-regions and the number of possible histories in each region is finite, every history must be repeated an infinite number of times in the universe.

**4. A critic to the Ellis-Brundrit (E-B) scenario**

We first discuss the E-B reasoning.

1. Step 1 in the E-B argumentation states that an infinite homogeneous universe will contain an infinity of galaxies and planets. This is correct. As these entities are countable, that infinity would be $\aleph$-0 (the number of natural numbers).

2. On the other hand, the number of history lines mentioned in step 2 is not countable. In a classic universe with a continuous non-quantified space, which they are considering, that number would be the number of lines that could be drawn in the infinite volume of the configuration space defined by all the independent parameters of the universe. That infinity is greater than $\aleph$-0, and also greater than the continuum.

3. Step 3 is correct.

4. To estimate the probability of the emergence of living beings, according to step 4, it would be necessary to estimate the number of history lines that make this emergence possible. Since the total number of lines is infinite (this is an E-B hypothesis), for that probability to be greater than zero, the number of histories compatible with life should be infinite.

    As E-B correctly indicate in a endnote, from our existence it cannot be deduced that the probability is greater than zero. We must add that the same would happen even if we had a finite number of positive cases of life in other planets. To achieve

an infinite number of histories, they assume that all the lines in a range around ours (which did produce life) in the configuration space would give rise to the same or a very similar result. If the radius of that range were greater than zero, the number of history lines inside the range would be infinite.

5. For the time being, we accept step 5.

6. If all the previous steps were correct, steps 6 and 7 (the E-B conclusion) would follow correctly from the premises. The 7.b quotation summarizes the conclusion. The 7.a quotation, however, is not valid. It should say just the opposite: *...with an infinite number of histories to look at, it is difficult to provide an argument why a particular history should occur more than once.* This is so, because the infinity of lines is much larger than the infinity of living beings (which is $\aleph$-0), therefore each planet compatible with life could be associated to a different history, even if their number were infinite.

In our opinion, the most dubious step in this development is step 4. Is it true that by considering almost-identical histories we attain a probability greater than zero of the emergence of DNA-based living beings, as E-B affirm?

This assertion can be refuted in two ways:

1. Refutation based on the probability that two beings/planets have very similar histories.

    a. A history line in our universe just now (ours, for instance) is a curve in the configuration space of the universe (which is infinite by hypothesis) with a length of 13,700 million years (since the big bang to our days).
    b. We will call a history line g *almost equal* to another history line h when all the points in g are located inside a cylinder with a finite radius E, built around the points of h.
    c. Consider two planets, A and B. Let h be the history line of A. Which is the probability that g (the history line of B) is *almost equal* to h?
    d. Although given h there is an infinity of g curves g that comply with the definition, that probability is the quotient the volume (in the configuration space) occupied by all the curves complying with the definition, divided by the volume occupied by all the curves. The former is the volume of a cylinder with radius E and a length of 13,700 million years, which is finite. The second is infinite (we have assumed that

the configuration space is infinite). Ergo the probability that two planets A and B have almost equal curves is zero. This does not preclude e.g. that there are some living planets apart from the Earth, but their number need not be infinite.

2. Refutation based on the chaotic character of the expansive universe.

    a. The uncertainty principle sets a minimum difference for the initial conditions of two different history lines.
    b. The equations of classical relativistic physics (those used by E-B) are chaotic.
    c. Therefore two different history lines, which at the beginning differ at least in the limits set by the uncertainty principle, must separate arbitrarily along their history, according with the definition of chaotic functions. This is even more true because the universe expands.
    d. Therefore there are no two *almost equal* history lines.

Would this argumentation be affected if the universe contained attractors where the history lines would converge? Those attractors could be:

a) Sinks (black holes). They are incompatible with life.

b) Cycles. They don't seem to exist in the universe. Perhaps they are incompatible with an expanding universe.

c) Saddles. They are unstable and sooner or later the history lines converging there would get apart in opposite directions.

d) Strange attractors. Those history lines attracted by them would get very near to the attractor, but not to one another, for given a point in one, the next point would have an unpredictable position. Ergo two history lines attracted by the same strange attractor would be very different, even though superficially similar.

We don't know if the universe contains strange attractors, they may be incompatible with its expansion. On the other hand, if there are, they may be incompatible with life. In that case, as the E-B argumentation refers to living being histories, they should be excluded.

> This argumentation is valid for sufficiently long sections of history lines, even though they do not start at the big bang.

In summary, the Ellis-Brundrit argumentation cannot stand as formulated in their original paper. This result is important, as it gainsays the intuitive and well-known idea that in an infinite universe all the histories must be repeated automatically, including those of living beings on the Earth[10]. Actually, the repetition of histories is avoided because the infinite of possible histories is larger than the infinite of objects (galaxies, planets) in an infinite universe. Therefore, for their argumentation about the repetition of the histories of living beings on the Earth to be valid, Ellis and Brundrit would need to prove that the number of histories leading to life is not concentrated in a finite volume in configuration space, or alternatively they would have to assert that the configuration space is finite. But then they would need to change their model of the universe, possibly going to a quantum model similar to that proposed by Garriga and Vilenkin.

### 5. A critic to the Garriga-Vilenkin (G-V) scenario

This argumentation is based on a set of currently questionable hypotheses. Therefore, their credibility is very low (because of the strong increase of uncertainty in a cascade of hypotheses). For instance:

1. It starts by assuming an infinite space, which is currently conjectural. We'll come back to this in the next section.

2. Space is assumed to be quantified and quantum theory is applied to cosmology. This is just another conjecture. Even more, all the attempts which have been done in past decades to find an indication of the quantum character of space and gravity have given a negative result[11].

---

[10] As a sample of the usual association between the infinite universe and the repetition of histories, we can mention the following paragraph of a recent book: «In an infinite Universe, every possible event might happen. In fact, in an infinite Universe every possible event does happen. Not just that: it happens an infinite number of times. While I am writing these lines, somewhere else, in an infinitely remote and unreachable point of the Cosmos, well out of our horizon, an exact replica of me on a planet perfectly identical to ours is writing the exact same words, but with a spelling error. In some far-off place, another me is doing the same thing, but he wears a red sweater rather than a blue one.» Balbi, A. (2008) The Music of the Big Bang. The Cosmic Microwave Background and the New Cosmology (Berlin, Springer) 138. Similar assertions can be found in texts by very many authors.

3. Their quantification is based in the DH interpretation of quantum mechanics. This interpretation has been the object of strong critics, and is far from being the standard interpretation among specialists[12].

4. The authors themselves accept that the fourth step in their argumentation (the inference that the total number of distinct histories is finite) is valid only in an universe without gravity (i.e. empty). In our universe there are situations (inside black holes or at the big bang) where the energy density of a single region may be infinite, but they reject those situations because they are far from the world of human experience. The original text says the following:

«In the absence of gravity, the boundedness of physical quantities usually follows from the finiteness of the energy of the system. The gravitational energy is not positive-definite,and in the presence of gravity the matter energy density can get arbitrarily large when it is compensated by equally large negative gravitational energy density. This happens, for example, during gravitational collapse in the interiors of black holes, in a recollapsing closed universe, and could happen in some extremely large quantum fluctuations. These extreme situations, however, are far removed from the low-energy world of the human experience, and it seems reasonable to isolate them when comparing the histories of different $O$-regions»[13].

However, this reasoning presents several problems:

a) It cannot be denied that, for the time being, human beings do not have much to do with black holes. That other intelligent beings or we ourselves in the future will be in the same situation seems an undue assumption. Perhaps one day we'll be able to take advantage of a black hole to reach relativistic speeds, in a similar way as the Voyager spacecrafts used Jupiter's gravity to accelerate themselves toward Saturn. In that case, we'd be extracting energy from the black hole, which would come to be a part of our history.

---

[11] For instance, it has been proposed that we could find an inkling of the quantum character of space by analyzing the arrival times of photons coming from remote cosmic explosions such as GRB [gamma ray bursts] and other high energy cosmic phenomena, studied by particle astrophysics. However, all the data currently available are consistent with the classic character of space. This is explained in more depth at Soler Gil, F. J. (2012), *Discovery or construction? Astroparticle physics and the search for physical reality* (Frankfurt, Peter Lang).

[12] An extensive critic of this interpretation can be found at Dowker, F - Kent, A (1996) «On the Consistent Histories Approach to Quantum Mechanics». Available at http://arxiv.org/abs/gr-qc/9412067.

[13] Garriga, J. - Vilenkin, A. (2001) 2-3.

b) But there is an additional important point that has not been considered in the G-V argumentation: the fact that all history lines did go once through a situation similar to the center of a black hole, namely the big bang. Ergo the conditions with infinite energy affect all the possible history lines. The number of distinct history lines may thus be infinite. As a result, the history lines have not to repeat themselves, even in an infinite universe.

c) On the other hand, the idea that the histories of habitable regions can be described ignoring gravitational singularities does not take into account that these singularities exist inside the $O$-regions and clearly influence their dynamics. For instance, our planet rotates around the black hole in the center of the Milky Way. Therefore, our own history depends to a certain extent of phenomena and regions related to a gravitational singularity. In summary, ignoring gravity while describing the $O$-regions does not seem feasible.

5. Finally, the fact that the universe expands may affect drastically the G-V argumentation. In a stationary universe, a finite number of histories could have no option but repeat. In a dynamic universe, however, the number of bins in each $O$-region may increase with time in such a way that histories get apart from one another indefinitely. This would increase the number of possible histories (perhaps without limit) thus preventing repetitions. Therefore the G-V argumentation cannot work without a proof that the number of bins relevant to building the histories is bounded in time in an expanding universe. And this is regardless of whether the gravitational singularities should be considered or not as part of the picture. G-V have not given any reason to suspect the existence of such a boundary, nor even signaled this difficulty.

Every one of the ideas postulated and the argumentative options in this list is doubtful. And the G-V argumentation requires all of them to be correct. Therefore, the scenario described by these authors must be considered highly speculative, to say the least.

**6. The uncertain character of an infinite universe**

The analysis of the Ellis-Brundrit and Garriga-Vilenkin argumentations has made it clear that the idea of a spatially infinite universe must be postulated at the very beginning of this type of conjectures. However, if the history of physics can be used to assess the initial credibility of a hypothesis, the postulate of a spatially infinite universe does not seem particularly promising.

Paraphrasing Aristotle, we can say that *nature abhors infinity*[14]. Along the history of physics, once and again situations have emerged where infinities seemed impossible to avoid. One after another they have fallen, and improved theories have avoided the infinites[15].

Just now we have two basic theories in physics, general relativity and quantum theory, which seem incompatible with one another. This makes it impossible to predict what happened before Planck's time, where both should be applied at the same time. However, these two theories have a weak point: both predict infinities: relativity, in the gravity singularities (black holes and the big bang); quantum theory, in the energy of the vacuum and those quantities that must be renormalized in quantum field theory[16]. The experience provided by the history of physics seems to indicate that both theories could be simple approximations of a third more general theory, which would unify them and eliminate the infinities. Perhaps, when this theory is found, the cosmologic infinities may also disappear.

However, while we wait for possible changes due to an important forward move in our comprehension of the physical world, it is usually claimed that the current standard cosmological model leaves open the possibility that the universe is infinite, since this is the consequence of Einstein's equations whenever K ≤ 0, where K defines the curvature of the universe in those equations.

---

[14] This idea is also spread among physicists. For instance, we can find it in the same Balbi book quoted above: «We are used to thinking that nothing in nature is really infinite: but we seem oddly inclined to accept that the Universe might be. However, nature abhors infinity, and with good reason.» Balbi, A. (2008) The Music of the Big Bang. The Cosmic Microwave Background and the New Cosmology» (Berlin, Springer) 138.

[15] The historian of physics, Michael Riordan, for instance, has mentioned Dirac's opinion in a recent article about this point: «In the early 1980s, Nobel laureate Paul Dirac told Princeton University theorist Ed Witten that the most important challenge in physics was "to get rid of infinity"» Riordan, M. (2012), «Tackling Infinity». Available at: http://www.americanscientist.org/bookshelf/pub/tackling-infinity.

[16] A recent description of the history of renormalization as a procedure to eliminate infinities in quantum field theory is Close, F. (2011), *The Infinity Puzzle: Quantum Field Theory and the Hunt for an Orderly Universe* (New York, Basica Books).

However, the infinitude of the universe can only be deduced in that case if an additional postulate is accepted, namely the so-called «cosmological principle», which asserts that the universe is homogeneous and isotropic at sufficiently large scales. This "principle" is needed to extrapolate the cosmological description of the universe beyond the boundaries of the observable universe. However, the justification of this principle gives rise to serious problems[17]. Therefore an agnostic position with respect to the extrapolations of the description of the universe beyond our horizon seems to be epistemologically reasonable, especially when that extrapolation goes to infinity.

In consequence, independently of the weak points in the Ellis-Brundrit and Garriga-Vilenkin argumentations we have signaled in the previous sections, the mere fact that both depend on the assumption that the universe is infinite gives them a highly speculative character.

## 7. Theoretical and practical conclusions

From the discussion of the Ellis-Brundrit and Garriga-Vilenkin argumentations in the preceding sections, it follows that the scenario of a universe where history repeats in an infinity of worlds spread throughout an infinite space cannot be considered a consequence of current physics and cosmology. In actual fact, these scenarios remain no more than literary tales, although they have been presented with the terminology of modern cosmology.

---

[17] A comprehensive discussion of the problems raised by the justification of the «cosmological principle» can be found in Beisbart, C. (2009), «Can we justifiably assume the Cosmological Principle in order to break model underdetermination in cosmology?» *Journal for General Philosophy of Science*, Volume 40, Issue 2 (2009), 175-205. And with regard to empirical data, new structures of increasing size are found now and again. [See, e.g. Clowes R. et al. (2012), «A structure in the early universe at z ~ 1.3 that exceeds the homogeneity scale of the R-W concordance cosmology». Available at: http://arxiv.org/abs/1211.6256]. Is there in fact a limit to the inhomogeneity of the universe? No one really knows.

This is, however, a theoretical conclusion. In practice, authors[18] and popular writers usually present these speculations as though they were very probable, practically proved by science. Even technical articles contain statements expressed with surprising certainty. For instance, in «Many worlds in one» Garriga and Vilenkin state:

«The argument below is far from being a rigorous proof of these statements, but we do believe that it makes them rather plausible»[19]

«Rather plausible»? This surprising certainty becomes temerity when we go from the technical articles to popularizing books. For instance, in a recent book of this character, with the same title as the article discussed here, Alex Vilenkin states:

«We arrive at the inevitable conclusion that every single history should be repeated an infinite number of times. [...] And any history that has a nonzero probability will happen —or has happened: in an infinite number of O-regions!»[20]

«Inevitable conclusion»? Perhaps the only inevitable conclusion of this is the fact that at least some of the supposedly scientific cosmological scenarios currently popularized are actually good samples of «ironic science», —using John Horgan's terminology—, that is:

«[...] opinions -which are at best "interesting" and provoke further comment. But it does not converge on the truth. It cannot achieve empirically verifiable "surprises" that force scientists to make substantial revisions in their basic description of reality»[21].

---

[18] We must exclude Ellis from the set of authors that act in this way. In fact, the article by Ellis and Brundrit analyzed here mention several factors that would invalidate their argumentation, and state clear reservations about the validity of the scenario they describe.

[19] Garriga, J. - Vilenkin, A. (2001) 1.

[20] Vilenkin, A. (2006), Many worlds in one (New York, Hill and Wang) 11-112.

[21] Horgan, J (1996), The end of science (Reading MA, Addison-Wesley) 7.